 \documentclass[aip,apl,reprint,showpacs,amsmath,amssymb]{revtex4-1}
\usepackage{natbib}
\usepackage{float}
\usepackage[dvips,final]{graphicx}
\usepackage{color}
\usepackage{amsmath}
\usepackage{mathrsfs} 

\usepackage{graphicx, subfigure}
\usepackage{pslatex}
\usepackage{relsize}
\usepackage{bm}
\usepackage{xspace} 

\usepackage{xcolor}
\usepackage{soul}

\def\ba{\begin{eqnarray}}
\def\ea{\end{eqnarray}}
\def\be{\begin{equation}}
\def\ee{\end{equation}}

\bibpunct{[}{]}{,}{n}{}{} 

\begin{document}

\title{Coupling of skyrmions mediated by RKKY interaction}

\author{R. Cacilhas}
\affiliation{Universidade Federal de Vi\c cosa, Departamento de F\'isica,
Av. Peter Henry Rolfs s/n, 36570-000, Vi\c cosa, MG, Brasil}
\author{S. Vojkovic}
\affiliation{Instituto de F\'isica, Pontificia Universidad Cat\'olica de Chile, Campus San Joaqu\'in Av. Vicu\~na Mackena,
4860 Santiago, Chile}
\author{V.L. Carvalho-Santos}
\affiliation{Universidade Federal de Vi\c cosa, Departamento de F\'isica,
Avenida Peter Henry Rolfs s/n, 36570-000, Vi\c cosa, MG, Brasil}
\author{E. B. Carvalho}
\affiliation{Universidade Federal do Vale do S\~ao Francisco, Campus Juazeiro, Av. Ant\^onio C. Magalh\~aes 510, 48902-300, Juazeiro, BA, Brasil}
\author{A.R. Pereira}
\affiliation{Universidade Federal de Vi\c cosa, Departamento de F\'isica,
Avenida Peter Henry Rolfs s/n, 36570-000, Vi\c cosa, MG, Brasil}
\author{D. Altbir}
\affiliation{Departamento de F\'isica, Universidad de Santiago de Chile and CEDENNA,\\ Avda. Ecuador 3493, Santiago, Chile}
\author{\'A. S. N\'u\~nez}
\affiliation{Departamento de F\'isica, Facultad de Ciencias F\'isicas y
Matem\'aticas, Universidad de Chile, Casilla 487-3, Santiago, Chile}
\date{\today}

\begin{abstract}

A discussion on the interaction between skyrmions in a bi-layer
system connected by a non-magnetic metal is presented. From
considering a free charge carrier model, we have shown that the
Ruderman-Kittel-Kasuya-Yosida ($RKKY$) interaction can induce
attractive or repulsive interaction between the skyrmions depending
on the spacer thickness. We have also shown that due to an
increasing in RKKY energy when the skyrmions are far from each
other, their widths are diminished. Finally, we have obtained
analytical solutions to the skyrmion position when the in-plane
distance between the skyrmions is small and it is shown that an
attractive RKKY interaction yields a skyrmion precessory motion.
This RKKY-induced coupling could be used as a skyrmion drag
mechanism to displace skyrmions in multilayers.

\end{abstract}

\maketitle

The possibility of using magnetic patterns such as vortices
\cite{Vortex}, domain walls \cite{Catalan} and skyrmions
\cite{Tomasello-SciRep-2014} in spintronic devices has resulted in
an increasing interest in studying statical and dynamical properties
of these magnetization collective modes. In particular, skyrmions
are topological spin textures that may appear as groundstate in
non-centrosymmetric crystals in the presence of the bulk
Dzyaloshinskii-Moriya interaction ($DMI$)
\cite{Jiang,Nagaosa-NatNano-2013,
Muhlbauer-Science-2009,Yu-NatMat-2011,Jonietz-Science-2010}. Due to
their topological stability, small size and low driving magnetic
field/current density \cite{Fert-NatNano-2013,Kang-SciRep-2016},
skyrmions are also promising candidates to compose spintronic
devices based on the interesting concept of racetrack memory
\cite{Tomasello-SciRep-2014,Parkin-Science-2008,Allwood-Science-2002},
as well as in logic devices \cite{Zhang-1,Zhang-2} and spin transfer
nano-oscillators \cite{Garcia-NJP-2016,Chui-AIP-2015}.

Nevertheless, as a consequence of the skyrmion Hall effect, the use
of these objects in racetrack devices is strongly hampered because a
skyrmion cannot move in a straight line along the driving current or
external magnetic field direction. Therefore, magnetic skyrmions can
be destroyed at the edges of
nanostripes\cite{Purkana-SciRep-2015,Zhang-3}. A possible way to
avoid the skyrmion Hall effect is the coupling between two skyrmions
lying in different layers. In fact, when two skyrmions are on
separate planes, changes in the interlayer exchange interaction and
the signs of the $DMI$ can induce different statical and dynamical
properties of the magnetization \cite{Koshibae-SciRep-2017}. In this
context, it has been shown that the Skyrmion Hall effect can be
suppressed by considering two perpendicularly magnetized
ferromagnetic sublayers strongly coupled via an antiferromagnetic
exchange interaction \cite{Zhang-4}. Furthermore, skyrmions
belonging to different layers can move simultaneously. That is, the
skyrmion in the layer without current follows the motion of the
skyrmion belonging to the layer in which an electrical current is
injected \cite{Zhang-4}. From experimental point of view, a
superimposition of skyrmions can be obtained from the strong dipolar
stray field of two skyrmions, which behave as a single particle
\cite{Sampaio-NatCom}.

In this paper, we study 
the statical and dynamical properties
of two skyrmions in superimposed layers connected by a non-magnetic
conductor material. The presence of the conductor causes the
magnetic layers interact through the
Ruderman-Kittel-Kasuya-Yosida($RKKY$)
interaction\cite{Kitel,Kasuya,Yosida,Bruno-PRB-46,Aristov-PRB}. The
$RKKY$ interaction is one of the most important and frequently
discussed couplings between the localized magnetic moments in solids
and adatoms
interactions\cite{Stepanyuk,Stepanyuk-2,Brovko,Ako-Nature}.
Particularly, concerning topological objects, this interaction has
been proposed as a mechanism to stabilize an isolated magnetic
skyrmion in a magnetic monolayer on a nonmagnetic conducting
substrate \cite{Bez-PRB}. Here, we show that due to the oscillatory
signal of $RKKY$ interaction as a function of the spacer thickness,
the interaction between skyrmions placed in different layers can be
attractive or repulsive. Additionally, we show that the skyrmion
radius diminishes when the skyrmions are far from each other,
recovering their widths of isolated skyrmions when they are
superimposed. Finally, we obtain an interaction potential for the
case of ferromagnetic $RKKY$ coupling and solve the Thiele's
equation aiming to describe the skyrmion dynamics when the in-plane
distance between the two skyrmions is small.

\begin{figure}
\includegraphics[scale=0.28]{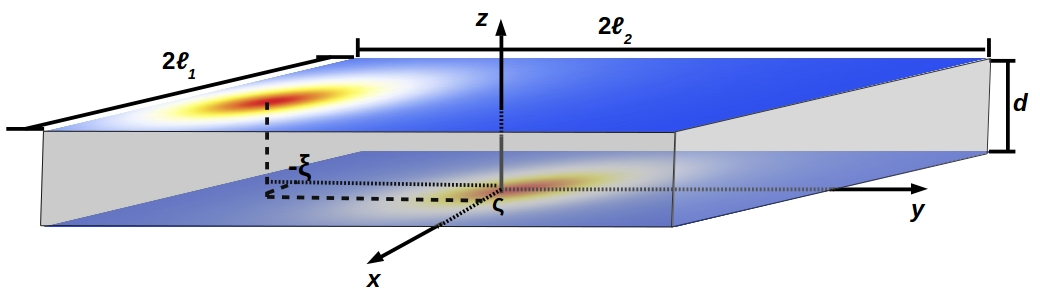}\hspace{0.1cm}
\includegraphics[scale=0.3]{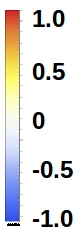}\caption{Adopted coordinate
system to describe the skyrmions position. Vertical bar shows the
$z$-component of the magnetization profile of the skyrmions. Gray
region represents the non-magnetic metal separating the
layers.}\label{Coordinates}
\end{figure}

The considered magnetic system consists of two rectangular monolayer
with dimensions $2\ell_1$ and $2\ell_2$ separated by a non-magnetic
metal with thickness $d$ (See Fig. \ref{Coordinates}). The layer
spacer consists of a conductor material and the electrons are
described as free charge carrier, whose formula to $RKKY$
interaction is well established \cite{Aristov-PRB,Karol-2008}.
Without lost of generality, it is assumed that the skyrmion placed
in the lower layer is located at the origin of the adopted
coordinate system in such way that the magnetization pattern of the
layers comprises a skyrmion positioned in the coordinate
$\mathbf{r}_1=(\zeta,-\xi,d)$ and a skyrmion at
$\mathbf{r}_2=(0,0,0)$, where the subindices $1$ and $2$ refer
respectively to the upper and lower layer.
\noindent The magnetic energy of an arbitrary magnetization profile
will be given by $E=E_{x}+E_{dmi}+E_{rkky}$, where $E_x$, $E_{dmi}$
e $E_{rkky}$ are respectively the energies coming from exchange,
Dzyaloshinskii-Moriya and $RKKY$ interactions. It is important to note that 
if the magnetic properties of the system is described only in terms of exchange and DMI 
interactions, helical states are predicted to  appear \cite{Nagaosa-PRB}. Nevertheless, despite anisotropy and Zeeman interactions are important to ensure skyrmion stability, for our purposes,
the knowledge on exchange and DMI energies is enough to describe our results. For more details and analysis of
another terms to the magnetic energy, see Ref. \cite{Aranda-JMMM}.

In the continuous
limit, the exchange and $DMI$ energies are respectively given by
\ba\label{ExModelDisc} E_x=J\int  (\vec{\nabla}\mathbf{m})^{2} dS\,,
\ea and \ba\label{DMIModelDisk}
E_{dmi}=D\int \left[ \left( -\mathbf{m}\times \frac{\partial\mathbf{m}}{\partial y}\right)\cdot\mathbf{x}
+ \left(\mathbf{m}\times \frac{\partial\mathbf{m}}{\partial x}\right)\cdot\mathbf{y}\right]      dS\,,\nonumber\\
\ea where $\mathbf{m}=\mathbf{M}/M_S$ is the unitary vector
describing the magnetization direction, $M_S$ is the saturation
magnetization of the material, $J$ is the exchange constant and $D$
is the $DMI$ constant. The $RKKY$ interaction will be determined
from a continuous approximation considering that the properties of
the non-magnetic material connecting the layers can be well
described by a free electron gas. In this context, the energy coming
from the $RKKY$ interaction between two magnetic moments is given by
$H_{rkky}=F(k_F\rho)\,\mathbf{m}_1\cdot\mathbf{m}_2$
\cite{Kitel,Bruno-PRB-46,Aristov-PRB}, where $\mathbf{m}_1$ is a
magnetic moment in layer $1$, $\mathbf{m}_2$ is a magnetic moment in
layer $2$, $F(k_F\rho)$ is the function that determines the coupling
between two magnetic moments belonging to different layers, $k_F$ is
the Fermi vector of the conductor material and
$\rho=\sqrt{(x_1-x_2)^2+(y_1-y_2)^2+d^2}$ is the distance between
two magnetic moments in different layers. For a free electron gas,
the function $F(k_F\rho)$ is given by \cite{Karol-2008} \ba
F(k_{F}R) = \mathcal{C}_3\left(\frac{2k_{F}\rho\cos(2k_{F}\rho) -
\sin(2k_{F}\rho) }{k_{F}^4\rho^4}\right)\,, \ea where
$\mathcal{C}_3={A_3^2\mathcal{M}k_F^4}/{8\pi h^2}$, $A_3$ is the
exchange interaction between electrons in the magnetic layer and
conduction electrons, $\mathcal{M}$ is the effective mass of the
conduction electrons and $h$ is the Planck constant. It can be
observed that, due to the periodicity of trigonometric functions,
the coupling between two magnetic moments can be ferromagnetic or
antiferromagnetic, depending on the distance between them. The total
$RKKY$ interaction energy is given by the sum of all pairs of
magnetic moments of the bi-layer. Thus, in a continuous
approximation, the $RKKY$ energy can be written as
\begin{equation}\label{RKKYModel}
E_{rkky}= \frac{1}{\mathcal{S}^2}\iint \mathbf{m}_{1}.\mathbf{m}_{2}'\,F(k_fR)\,    dS_{1}dS_{2}'\,,
\end{equation}
where $\mathcal{S}$ is the surface area of a unitary cell of the magnetic material
and the integrals are performed along the surfaces of the two magnetic
layers. It is important to note that the interaction is inversely proportional
to $\rho^4$ and for $\rho\gg \sqrt{a}$, $F(k_F\rho)$ oscillates with a period
$T = \Lambda_f/2$ and decays as $\rho^3\,\,$ \cite{Bruno-PRB-46}.

In the adopted model, the magnetization vector field
$\mathbf{M}(\mathbf{r})$ is considered as a continuous function
depending on the position inside the magnetic layer. There are
several ansatz describing the magnetization profile of a skyrmion
\cite{Beg-Arxiv,Mourafis-PRB-2006,Finazzi-PRL-2013,Vagson-JMMM-2015,Vagson-JAP-2015}.
In this work, we use the ansatz of Refs.
\cite{Guslienko-IEEEMagLett-2015,Aranda-JMMM}, in which the
magnetization is parametrized as
$\mathbf{m}_j=\sin\Theta_j\cos\Phi_j\,\mathbf{x}+\sin\Theta_j\sin\Phi_j\,\mathbf{y}+\cos\Theta_j\,\mathbf{z}$,
with \ba\label{ansatz}
\Theta_j=\arccos\left(\frac{\lambda^2-\mathcal{R}_j^2}{\lambda^2+\mathcal{R}_j^2}\right),\hspace{0.3cm}\Phi=\arctan\left(\frac{y-y_j}{x-x_j}\right),
\ea where  $\lambda$ is the skyrmion characteristic length and
$\mathcal{R}_j=\sqrt{(x-x_j)^2+(y-y_j)^2}$. The subindex $j=(1,2)$
describes the layer in which the skyrmion $j$ lies, that is,
$(x_j,y_j)=(\zeta,-\xi)$ for layer $1$ and $(x_j,y_j)=(0,0)$ for
layer $2$. Therefore, Eq.(\ref{ExModelDisc}) can be rewritten as
\ba\label{ExModel} E_{x}=
J\int\sum_{i=1,2}\left[\left(\frac{\partial\Theta_j}{\partial
x_i}\right)^2+\sin^2\Theta_j\left(\frac{\partial\Phi_j}{\partial
x_i}\right)^2\right]dS_j\,. \ea Subindex $i$ in the sum refers to
$x$ (1) and $y$ (2) coordinates. In addition, Eq.
(\ref{DMIModelDisk}) is also rewritten as \ba\label{DMIModel}
E_{dmi}= D \int \left[ \sin\Phi_j \frac{\partial\Theta_j}{\partial
y} + \cos\Phi_j \frac{\partial\Theta_j}{\partial x}
\right.\nonumber\\\left.+\frac{\sin2\Theta_j}{2}
\left(\cos\Phi_j\frac{\partial\Phi_j}{\partial y}
-\sin\Phi_j\frac{\partial\Phi_j}{\partial x} \right) \right] dS_j\,.
\ea

\begin{figure}
\includegraphics[scale=0.25]{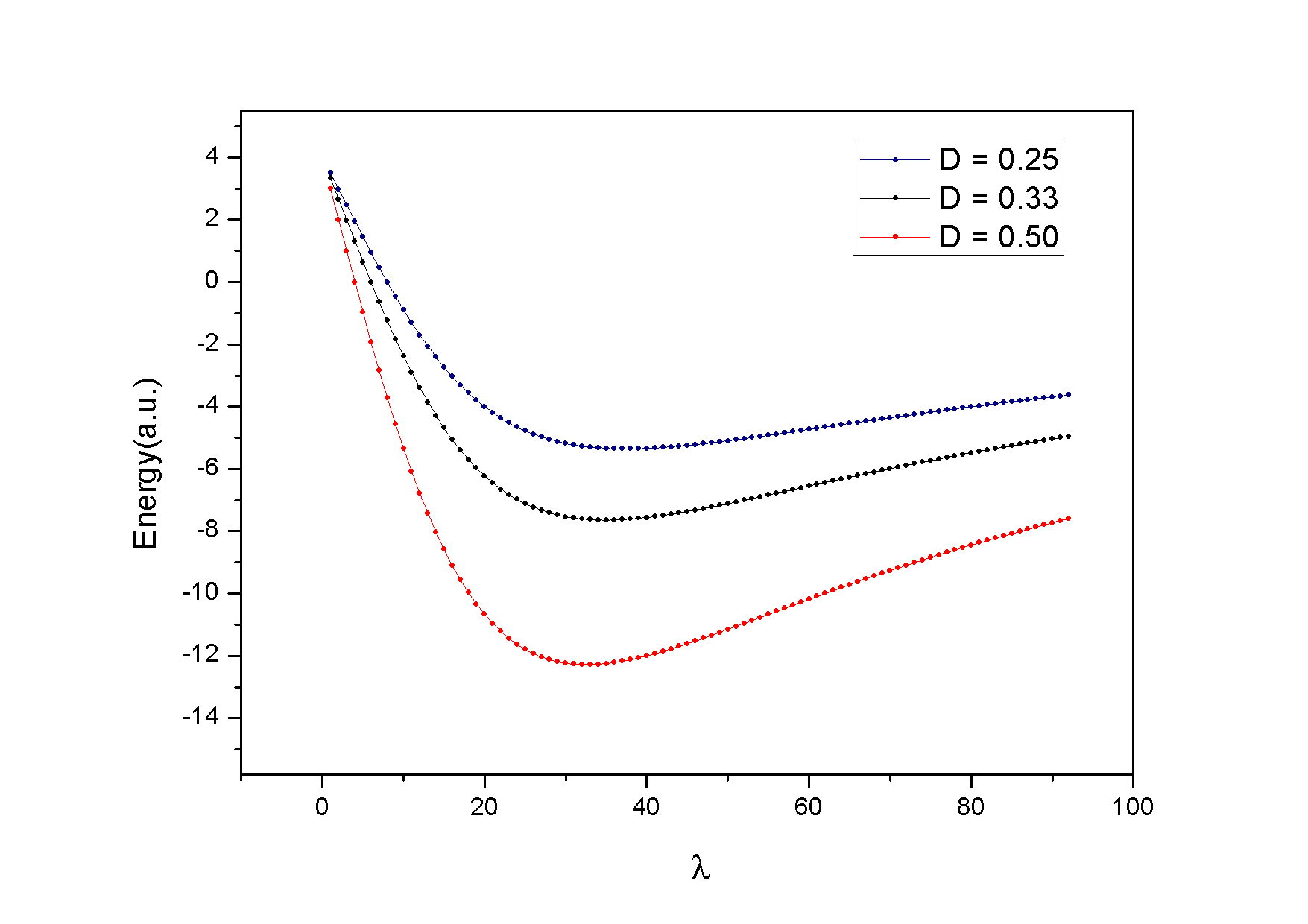}\caption{Qualitative description
of $E_x+E_{dmi}$ energy given as a function of $\lambda$ for
different relations $J/D$. In this case, we have used $J=1$ and
$D=J/4$ (blue), $D=J/3$ (black) and $D=J/2$ (red).}\label{SkWidth}
\end{figure}

From the described model we are in position to calculate the total
magnetic energy of the bi-layer system. Therefore, from considering
that the skyrmions are far from the borders of the stripe, the
exchange and $DMI$ energies of the described skyrmion profile are
given respectively by $ E_{x_j}=4\pi\lambda^2
J\mathcal{G}_j$ and $E_{dmi_j} =
2\pi\lambda^3D\mathcal{G}_j$, where
$$\mathcal{G}_j=\frac{1}{\mathcal{R}_j^2+\lambda^2}-\frac{1}{R_j^2+\lambda^2},$$$$R_j=\text{min}[(L_1-x_j)^2+(L_1-y_j)^2,(L_2-x_j)^2+(L_2-y_j)^2].$$
It can be noted that, in the limit $R_j\gg\lambda$, the exchange
energy of the skyrmions is $E_x\approx8\pi J$ and $E_{dmi}\approx4\pi\lambda D$, 
which is in accord to
the energy of solitonic solutions of the non-linear $\sigma$-model
\cite{Rajaraman}. Figure \ref{SkWidth} shows the skyrmion energy as a function of its width
for different $J/D$. It can be noted that, despite the energy
decreases when $D$ increases, there is no qualitative changes of the $\lambda$ value 
that minimizes the magnetic energy.
This is associated with the fact that we are not considering
anisotropy or Zeeman interactions in this work. For more details and analysis on the skyrmion size, 
see Ref. \cite{Wang-Comm-Phys}.


\begin{figure}
\includegraphics[scale=0.25]{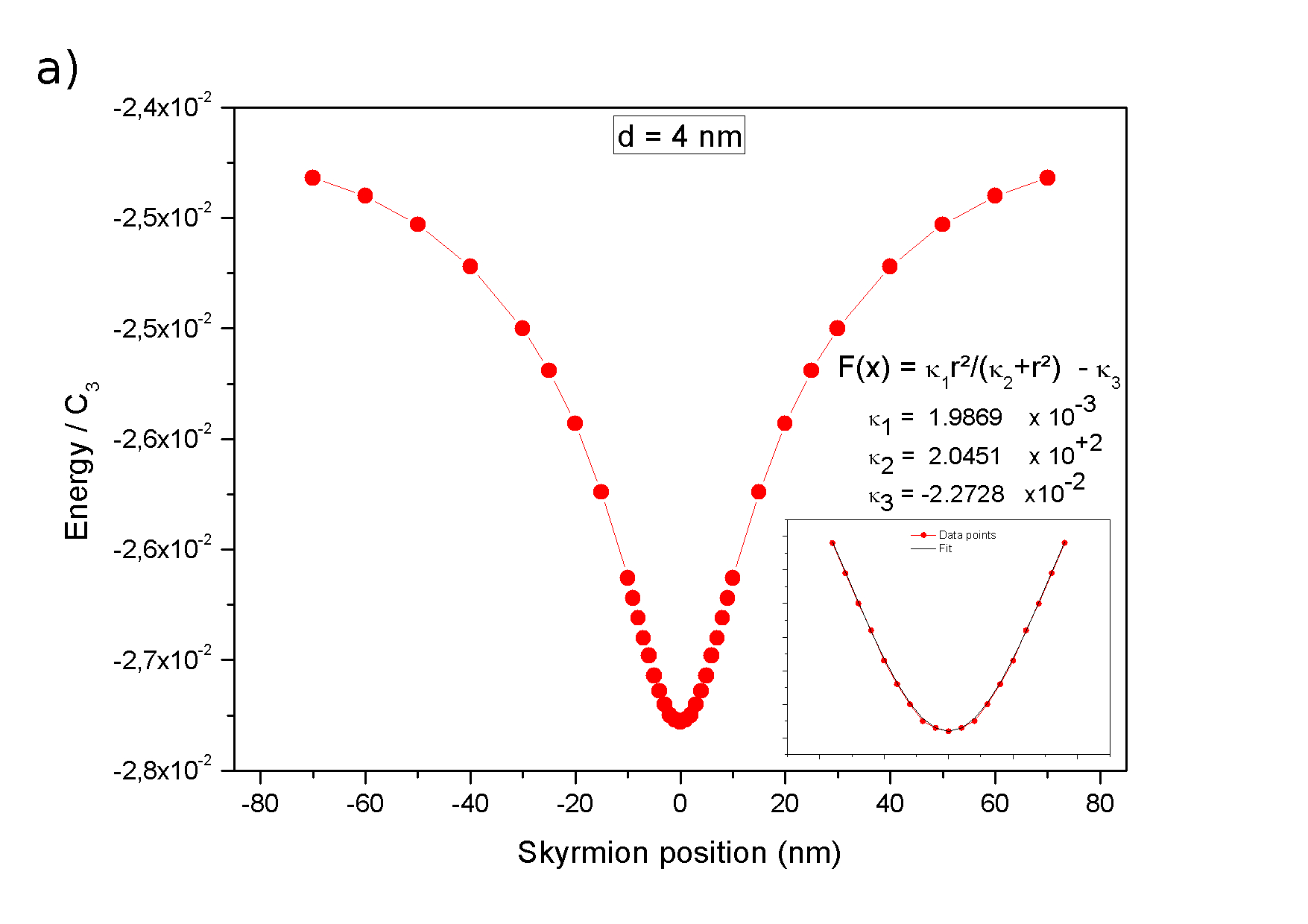}\\\includegraphics[scale=0.25]{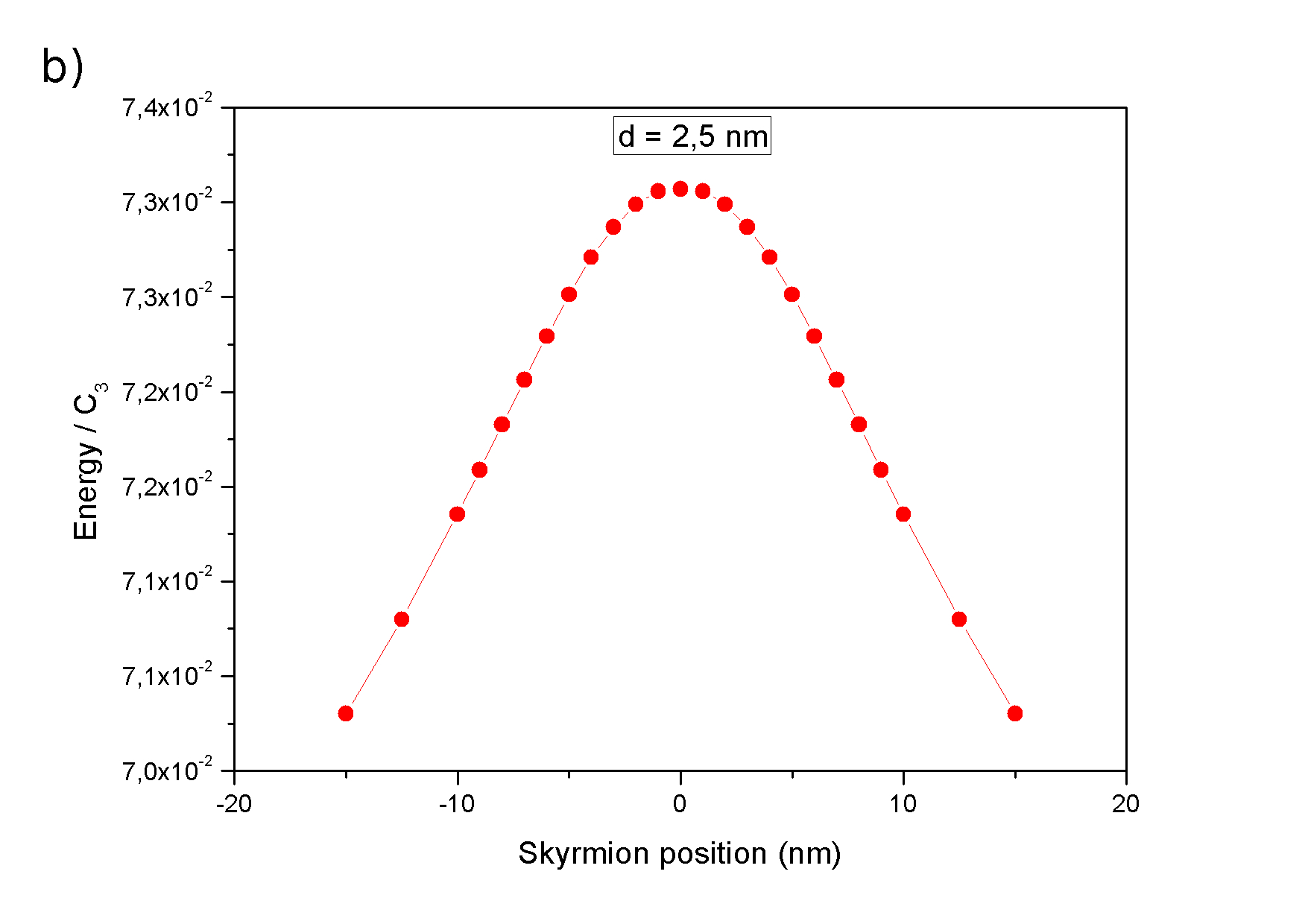}\caption{$RKKY$ energy
between skyrmions as a function of the interlayer distance. Figure
(a) shows the $RKKY$ energy as a function of the distance between
the skyrmions for an interlayer distance of $4$ nm. Inset is a view
of the region in which the in-plene distance between skyrmions is
less than $10$ nm. Fig. (b) presents $RKKY$ interaction when the
interlayer distance is $2.5$ nm. }\label{RKKYEnergy}
\end{figure}

\begin{figure}
\includegraphics[scale=0.13]{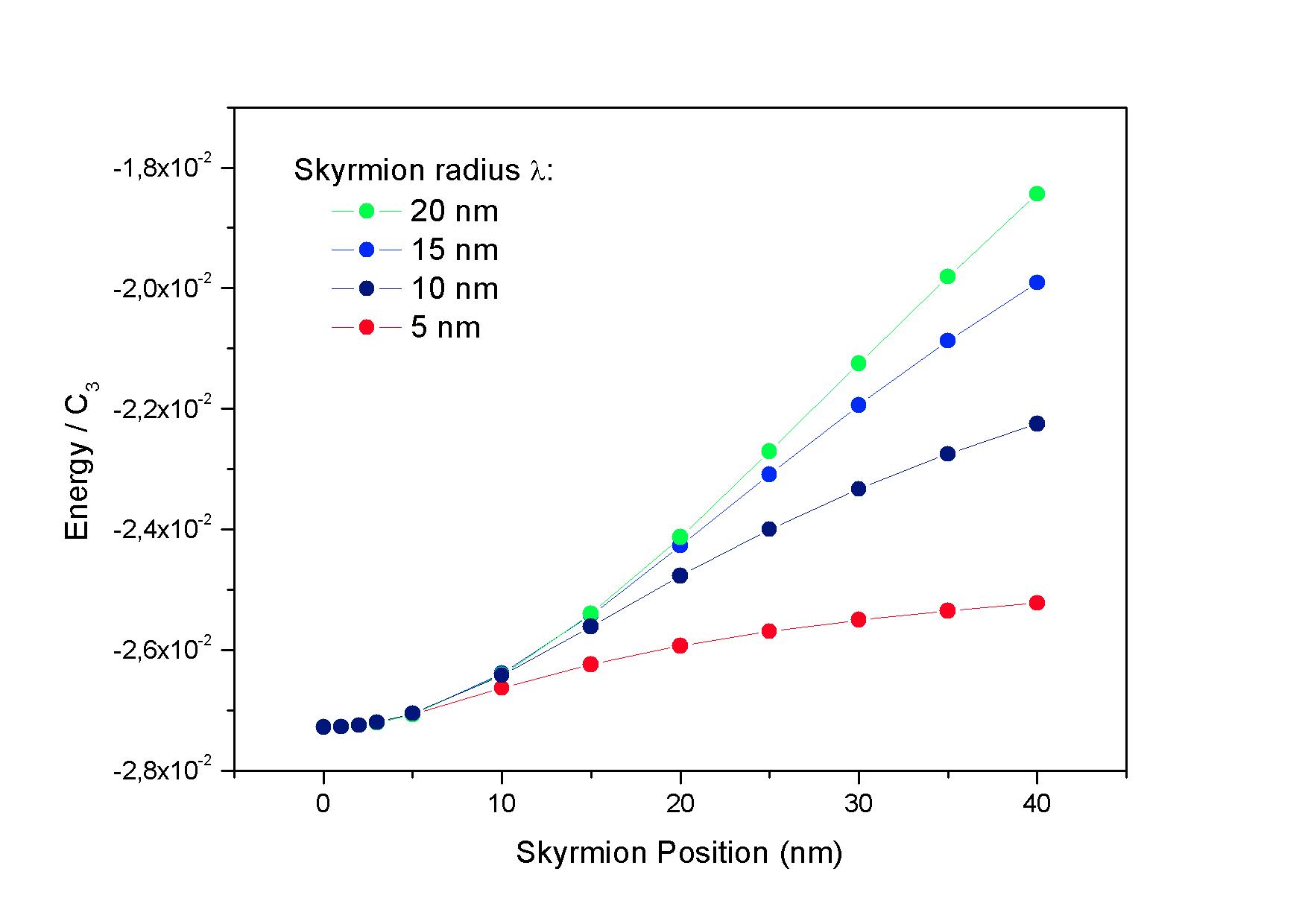}\caption{$RKKY$ energy for
different values of skyrmion widths for $d=4$ nm. It can be noted
that the skyrmions energy increases with $\lambda$ for large
distances. However, when the distance between the skyrmions is
$\approx d$, the $RKKY$ energy is independent on the skyrmions
width.}\label{SkLenght}
\end{figure}

Our main goal in this work is to determine the magnetic energy when
the skyrmions are separated by a conductor layer. Nevertheless, an
expression for $RKKY$ energy is hard to be obtained analytically and
so, it is numerically solved by the integration of
Eq.(\ref{RKKYModel}). We consider that the conductor material is a
copper layer for which $k_f=1.36\times10^{10}m^{-1}$. The
integration along all the region of the stripe demands a very long
computational time. Therefore, aiming to obtain numerical results
faster, we have performed the numerical integration by cutting the
integration region. That is, we have calculated the $RKKY$ energy of
one magnetic moment located in layer $1$ with the magnetic moments
located inside a square of side $\rho$ in the layer $2$. This
procedure was performed for all magnetic moments in layer $1$. It is
noted that differences in the obtained $RKKY$ energy values starts
to be negligible when the square side is on the order of $1.2$ nm.
Thus, by using this value to the square side, we have calculated the
$RKKY$ energy as a function of the spacer thickness and the skyrmion
position along $y$-direction. The main results are summarized in
Fig.\ref{RKKYEnergy}, in which it can be noted that due to the
periodicity of the $RKKY$ interaction, the skyrmion-skyrmion
interaction can be attractive or repulsive, depending on the
interlayer thickness. We have also performed the numerical
integration of Eq.(\ref{RKKYModel}) for different values of
$\lambda$ aiming to understand how $RKKY$ interaction influences the
skyrmion width; the results are shown in Fig.\ref{SkLenght}. We have
obtained that the $RKKY$ energy depends on the skyrmon width only
when they are far away one to another. That is, for large distances,
the skyrmion radius diminishes aiming to reduce $RKKY$ energy.
However, when the distance between them decreases, $RKKY$ energy is
independent of the skyrmion width and then, when the skyrmions are
superimposed, their widths must be determined by the interplay among
uniaxial anisotropy, exchange and $DMI$ interactions, being equal to the
isolated skyrmion width  \cite{Aranda-JMMM,Wang-Comm-Phys}.

\begin{figure}
\includegraphics[scale=0.2]{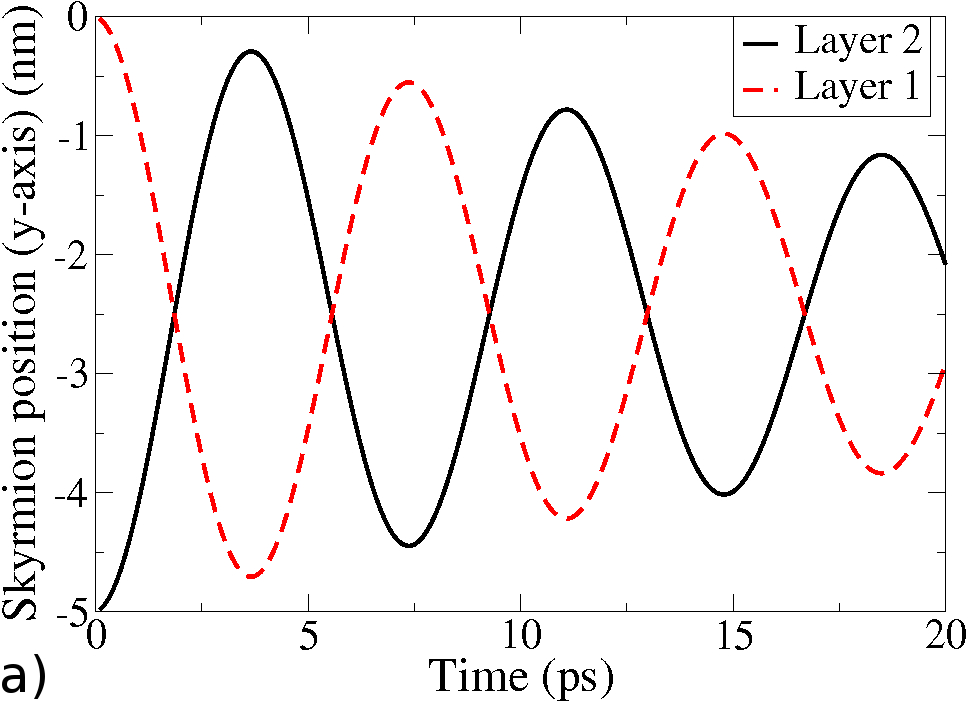}\\\includegraphics[scale=0.2]{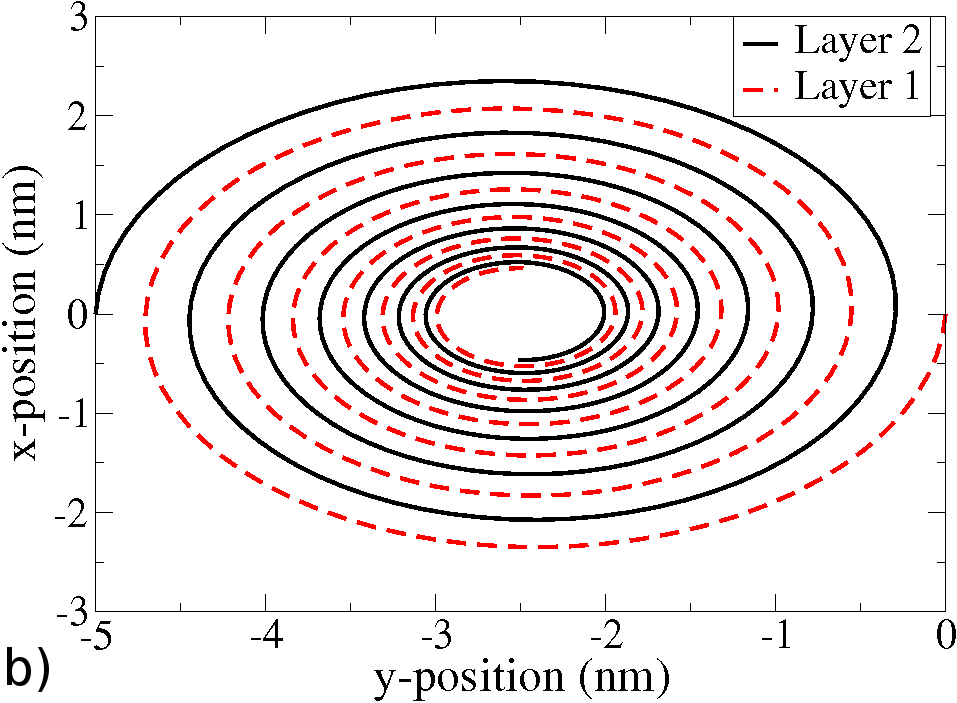}\caption{Fig.a shows
the positions along y-axis of skyrmions as a function of time. Fig.b shows the
precessory motion of both skyrmions around each other in the in-plane perspective. }\label{SkPosition}
\end{figure}

We will now describe the skyrmion dynamics from considering two
layers in the absence of currents and external magnetic fields. Additionaly,
we will assume that the in-plane skymion distance is on the
order of $\lambda$ because, for larger distances, the potential
coming from $RKKY$ interaction is practically constant. In our
analytical model, we neglect the dynamical deformation of the
skyrmions in such way that, the Landau-Lifshitz-Gilbert equation can
be reduced to the Thiele equation\cite{Thiele-Work}, written as
\ba\label{ThieleEq} -g\hat{z}\times\mathbf{v}_j(t)+\alpha
\mathcal{D}\mathbf{v}_j(t)=-\nabla U(\mathbf{r}_1-\mathbf{r}_2)\,,
\ea where subscripts $j$ label the layers ($1$ and $2$). The first
term in the above equation describes the Magnus force exerted by the
magnetic texture in the magnetic skyrmion located in the $j$-th
layer, which displaces with velocity $\mathbf{v}_j$. Once we are
considering the dynamics of a highly symmetrical skyrmion, we have
that $g=-4\pi M_s/\gamma$, where $\gamma$ is the gyromagnetic ratio.
The second term represents the dissipative force action in each
magnetic skyrmion. The right side of Eq.\eqref{ThieleEq} represents
the force that determines the skyrmions dynamics, which contains a
contribution from the potential $U(\mathbf{r}_1-\mathbf{r}_2)$,
originated from the skyrmion-skyrmion interaction. From the results
obtained to the $RKKY$ energy, we can state that when
$r=|\mathbf{r}_1-\mathbf{r}_2|$ is very small, the interaction
energy between skyrmions is given by a harmonic potential
$U_{r\rightarrow0}\approx\kappa_1 r^2$, where $\kappa_1$ depends on
the layer distance and the conductor parameters and $r$ is the
distance between the skyrmions along the $xy$-plane. From Fig.
\ref{RKKYEnergy} one can note that this approximation is very good
when the skyrmion centers are separated by a distance $r\lesssim20$
nm. On the other hand, in the asymptotic limit $r\gtrsim\lambda$,
the interaction energy is almost constant and comes from the
interaction of the skyrmion in layer $1$ with the ferromagnetic
state in layer $2$ and vice-versa, with
$U_{r\gtrsim\lambda}=\kappa_3$. In these asymptotic limits, we can
obtain the analytical solution to Eq.(\ref{ThieleEq}) and it is
presented into the Appendix \ref{Appendix}. By using the result
given in Fig.\ref{RKKYEnergy}, we can also assume that the potential
in the intermediary zone (that is, $20\,\text{nm}<r<70\,\text{nm}$)
can be very well represented by $U=\kappa_1
r^2/(\kappa_2+r^2)+\kappa_3$ (See Fig. \ref{RKKYEnergy}a). From estimating $\mathcal{C}_3\mathcal{S}^2\sim10^{15}$ Jm, we have obtained the position of the skyrmions for $d=4$ nm as a function of time and the results can be viewed in Fig. \ref{SkPosition}. It can be noted that
when the in-plane distance between skyrmions is on the order of
$\lambda$, a drag effect is observed and an attractive interaction is observed in such way that the two skyrmions become
coupled, precessing one around other with frequency $\omega={4\kappa_1 tg}/{(g^2+\alpha^2)}$. This precessory motion is resulted from the interplay between skyrmion-skyrmion attraction and the Magnus force exerted by the
magnetic texture in the magnetic skyrmions. In this way, when the skyrmions move one towards the other, the skyrmion Hall effect produces a displacement of skyrmions in opposite directions and the resulting motion is that one shown in Fig. \ref{SkPosition}b. The skyrmions can be decoupled
by changing the interlayer distance when an
antiferromagnetic $RKKY$ interaction appears. 

The observed skyrmion coupling is a very interesting result because 
it opens the possibility to realize devices based on the concept of skyrmion drag effects. 
That is, if it is possible to apply
different in-plane magnetic fields or different anisotropies in layers $1$ and $2$, the skyrmions can
displace with different velocities; however, when the distance
between them is small, skyrmions couple and can displace together, 
which can diminishes the skyrmion Hall effect due to the increasing of the effective skyrmion mass. 
In addition, a magnetoresistive device can be proposed because there is a
decreasing in the resistence when the skyrmion are superimposed.
In fact, the value of an electric current passing through the multilayer system 
in the direction perpendicular to the planes depends on the skyrmion positions because 
magnetoresistive effects take place and the resistence diminishes 
when the skyrmions are superimposed.

In conclusion, we have described the coupling between skyrmions in
different layers separated by a conductor non-magnetic material. It
has been shown that the $RKKY$ interaction can lead to a skyrmion
coupling/decoupling, depending on the spacer thickness. The skyrmion
radius is also affected by $RKKY$ interaction in such way that it
diminishes when the skyrmions are far from each other and it
increases when they are in an small in-plane distance. The dynamics
of this system show that when skyrmions in different layers are put
near one to another, they execute a precession motion around
each other. The drag effect mediated by $RKKY$ can be used to move
skyrmions in different layers with the same velocity.



\appendix

\section{Solution of Thiele Equation for interacting skyrmions}\label{Appendix}

The dynamics of the skyrmion pair is determined from the solution of Thiele's equation (\ref{ThieleEq}). In the limit of $r\rightarrow0$ the potential can be approximated by a harmonic oscillator of the kind $U=\kappa_1 r^2$. The analytical solution to the Thiele equation is given by
\begin{widetext}
\ba
x_{1_{r\rightarrow0}}=\frac{1}{2}\left\{\zeta+\text{e}^{\frac{-4\kappa_1 t \alpha}{16\pi^2+\alpha^2}}\left[\zeta\cos\left(\frac{16\kappa_1\pi t}{16\pi^2+\alpha^2}\right)-\xi\sin\left(\frac{16\kappa_1\pi t}{16\pi^2+\alpha^2}\right)\right]\right\}
\ea

\ba
y_{1_{r\rightarrow0}}=\frac{1}{2}\left\{-\xi-\text{e}^{\frac{-4\kappa_1 t \alpha}{16\pi^2+\alpha^2}}\left[\zeta\sin\left(\frac{16\kappa_1\pi t}{16\pi^2+\alpha^2}\right)+\xi\cos\left(\frac{16\kappa_1\pi t}{16\pi^2+\alpha^2}\right)\right]\right\}
\ea

\ba
x_{2_{r\rightarrow0}}=\frac{1}{2}\left\{\zeta+\text{e}^{\frac{-4\kappa_1 t \alpha}{16\pi^2+\alpha^2}}\left[-\zeta\cos\left(\frac{16\kappa_1\pi t}{16\pi^2+\alpha^2}\right)+\xi\sin\left(\frac{16\kappa_1\pi t}{16\pi^2+\alpha^2}\right)\right]\right\}
\ea
\ba
y_{2_{r\rightarrow0}}=\frac{1}{2}\left\{-\xi+\text{e}^{\frac{-4\kappa_1 t \alpha}{16\pi^2+\alpha^2}}\left[\zeta\sin\left(\frac{16\kappa_1\pi t}{16\pi^2+\alpha^2}\right)+\xi\cos\left(\frac{16\kappa_1\pi t}{16\pi^2+\alpha^2}\right)\right]\right\}
\ea
\end{widetext}



\end{document}